# Analysis of Empirical Software Effort Estimation Models

Saleem Basha
Department of Computer Science
Pondicherry University
Puducherry, India
smartsaleem1979@gmail.com

Dhavachelvan P
Department of Computer Science
Pondicherry University
Puducherry, India
dhavachelvan@gmail.com

*Abstract* – Reliable effort estimation remains an ongoing challenge to software engineers. Accurate effort estimation is the state of art of software engineering, effort estimation of software is the preliminary phase between the client and the business enterprise. The relationship between the client and the business enterprise begins with the estimation of the software. The credibility of the client to the business enterprise increases with the accurate estimation. Effort estimation often requires generalizing from a small number of historical projects. Generalization from such limited experience is an inherently under constrained problem. Accurate estimation is a complex process because it can be visualized as software effort prediction, as the term indicates prediction never becomes an actual. This work follows the basics of the empirical software effort estimation models. The goal of this paper is to study the empirical software effort estimation. The primary conclusion is that no single technique is best for all situations, and that a careful comparison of the results of several approaches is most likely to produce realistic estimates.

## I. INTRODUCTION

Software effort estimation is one of the most critical and complex, but an inevitable activity in the software development processes. Over the last three decades, a growing trend has been observed in using variety of software effort estimation models in diversified software development processes. Along with this tremendous growth, it is also realized the essentiality of all these models in estimating the software development costs and preparing the schedules more quickly and easily in the anticipated environments. Although a great amount of research time, and money have been devoted to improving accuracy of the various estimation models, due to the inherent uncertainty in software development projects as like complex and dynamic interaction factors, intrinsic software complexity, pressure on standardization and lack of software data, it is unrealistic to expect very accurate effort estimation of software development processes [1]. Though there is no proof on software cost estimation models to perform consistently accurate within 25% of the actual cost and 75% of the time [30], still the available cost estimation models extending their support for intended activities to the possible extents. The accuracy of the individual models decides their applicability in the projected environments, whereas the accuracy can be defined based on understanding the calibration of the software data. Since the precision and reliability of the effort estimation is very important for the competitiveness of software companies, the enterprises and researchers have put their maximum effort to develop the accurate models to estimate effort near to accurate levels. There are many estimation models have been proposed and can be categorized based on their basic formulation schemes; estimation by expert [5], analogy based estimation schemes [6], algorithmic methods including empirical methods [7], rule induction methods [8], artificial neural network based approaches [9] [17] [18], Bayesian network approaches [19], decision tree based methods [21] and fuzzy logic based estimation schemes [10] [20].

Among these diversified models, empirical estimation models are found to be possibly accurate compared to other estimation schemes and COCOMO, SLIM, SEER-SEM and FP analysis schemes are popular in practice in the empirical category [24] [25]. In case of empirical estimation models, the estimation parameters are commonly derived from empirical data that are usually collected from various sources of historical or passed projects. Accurate effort and cost estimation of software applications continues to be a critical issue for software project managers [23]. There are many introductions, modifications and updates on empirical estimation models. A common modification among most of the models is to increase the number of input parameters and to assign appropriate values to them. Though some models have been inundated with more number of inputs and output features and thereby the complexity of the estimation schemes is increased, but also the accuracy of these models has shown with little improvement. Although they are diversified, they are not generalized well for all types of environments [13]. Hence there is no silver bullet estimation scheme for different environments and the available models are environment specific.

## II. COCOMO ESTIMATION MODEL

### A. *COCOMO 81*

COCOMO 81 (Constructive Cost Model) is an empirical estimation scheme proposed in 1981 [29] as a model for estimating effort, cost, and schedule for software projects. It was derived from the large data sets from 63 software projects ranging in size from 2,000 to 100,000 lines of code, and programming languages ranging from assembly to PL/I. These data were analyzed to discover a set of formulae that were the best fit to the observations. These formulae link the size of the system and Effort Multipliers (EM) to find the effort to develop a software system. In COCOMO 81, effort is expressed as Person Months (PM) and it can be calculated as

$$PM = a * Size^b * \prod_{i=1}^{15} EM_i \quad (1)$$

where,

"a" and "b" are the domain constants in the model. It contains 15 effort multipliers. This estimation scheme accounts the experience and data of the past projects, which is extremely complex to understand and apply the same.

Cost drives have a rating level that expresses the impact of the driver on development effort, PM. These rating can range from Extra Low to Extra High. For the purpose of quantitative analysis, each rating level of each cost driver has a weight associated with it. The weight is called Effort Multiplier. The average EM assigned to a cost driver is 1.0 and the rating level associated with that weight is called Nominal.

### B. COCOMO II

In 1997, an enhanced scheme for estimating the effort for software development activities, which is called as COCOMO II. In COCOMO II, the effort requirement can be calculated as

$$PM = a * Size^E * \prod_{i=1}^{17} EM_i \quad (2)$$

where $\quad E = B + 0.01 * \sum_{j=1}^{5} SF_j$

COCOMO II is associated with 31 factors; LOC measure as the estimation variable, 17 cost drives, 5 scale factors, 3 adaptation percentage of modification, 3 adaptation cost drives and requirements & volatility. Cost drives are used to capture characteristics of the software development that affect the effort to complete the project.

COCOMO II used 31 parameters to predict effort and time [11] [12] and this larger number of parameters resulted in having strong co-linearity and highly variable prediction accuracy. Besides these meritorious claims, COCOMO II estimation schemes are having some disadvantages. The underlying concepts and ideas are not publicly defined and the model has been provided as a black box to the users [26]. This model uses LOC (Lines of Code) as one of the estimation variables, whereas Fenton et. al [27] explored the shortfalls of the LOC measure as an estimation variable. The COCOMO also uses FP (Function Point) as one of the estimation variables, which is highly dependent on development the uncertainty at the input level of the COCOMO yields uncertainty at the output, which leads to gross estimation error in the effort estimation [33]. Irrespective of these drawbacks, COCOMO II models are still influencing in the effort estimation activities due to their better accuracy compared to other estimation schemes.

### III. SEER-SEM ESTIMATION MODEL

SEER (System Evaluation and Estimation of Resources) is a proprietary model owned by Galorath Associates, Inc. In 1988, Galorath Incorporated began work on the initial version of SEER-SEM which resulted in an initial solution of 22,000 lines of code. SEER (SEER-SEM) is an algorithmic project management software application designed specifically to estimate, plan and monitor the effort and resources required for any type of software development and/or maintenance project. SEER, which comes from the noun, referring to one having the ability to foresee the future, relies on parametric algorithms, knowledge bases, simulation-based probability, and historical precedents to allow project managers, engineers, and cost analysts to accurately estimate a project's cost schedule, risk and effort before the project is started. Galorath chose Windows due to the ability to provide a more graphical user environment, allowing more robust management tradeoffs and understanding of what drives software projects.[4]

This model is based upon the initial work of Dr. Randall Jensen. The mathematical equations used in SEER are not available to the public, but the writings of Dr. Jensen make the basic equations available for review. The basic equation, Dr. Jensen calls it the "software equation" is:

$$S_e = C_{te}(Kt_d)^{0.5} \quad (3)$$

where,

'S' is the effective lines of code, 'ct' is the effective developer technology constant, 'k' is the total life cycle cost in man-years, and 'td' is the development time in years.

This equation relates the effective size of the system and the technology being applied by the developer to the implementation of the system. The technology factor is used to calibrate the model to a particular environment. This factor considers two aspects of the production technology -- technical and environmental. The technical aspects include those dealing with the basic development capability: Organization capabilities, experience of the developers, development practices and tools etc. The environmental aspects address the specific software target environment: CPU time constraints, system reliability, real-time operation, etc.

The SEER-SEM developers have taken the approach to include over 30 input parameters, including the ability to run Monte Carlo simulation to compensate for risk [2]. Development modes covered include object oriented, reuse, COTS, spiral, waterfall, prototype and incremental development. Languages covered are 3rd and 4th generation languages (C++, FORTRAN, COBOL, Ada, etc.), as well as application generators. It allows staff capability, required design and process standards, and levels of acceptable development risk to be input as constraints [15]. Figure 1 is adapted from a Galorath illustration and shows gross categories of model inputs and outputs, but each of these represents dozens of specific input and output possibilities and parameters.

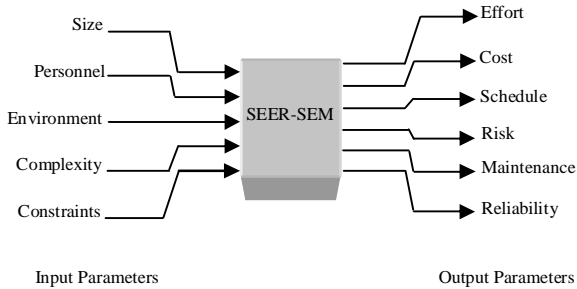

Figure 1. SEER-SEM I/O Parameters

Features of the model include the following:

- Allows probability level of estimates, staffing and schedule constraints to be input as independent variables.
- Facilitates extensive sensitivity and trade-off analyses on model input parameters.
- Organizes project elements into work breakdown structures for convenient planning and control.
- Displays project cost drivers.
- Allows the interactive scheduling of project elements on Gantt charts.
- Builds estimates upon a sizable knowledge base of existing projects.

Model specifications include these:

- Parameters: size, personnel, complexity, environment and constraints - each with many individual parameters; knowledge base categories for platform & application, development & acquisition method, applicable standards, plus a user customizable knowledge base.
- Predictions: effort, schedule, staffing, defects and cost estimates; estimates can be schedule or effort driven; constraints can be specified on schedule and staffing.
- Risk Analysis: sensitivity analysis available on all least/likely/most values of output parameters; probability settings for individual WBS elements adjustable, allowing for sorting of estimates by degree of WBS element criticality.
- Sizing Methods: function points, both IFPUG sanctioned plus an augmented set; lines of code, both new and existing.

- Outputs and Interfaces: many capability metrics, plus hundreds of reports and charts; trade-off analyses with side-byside comparison of alternatives; integration with other Windows applications plus user customizable interfaces.

Aside from SEER-SEM, Galorath, Inc. offers a suite of many tools addressing hardware as well as software concerns. One of particular interest to software estimators might be SEER-SEM, a tool designed to perform sizing of software projects.

The study done by Thibodeau in 1981 and a study done by IIT Research Institute (IITRI) in 1989 states that they calibrated SEER-SEM model using three databases. The significance of this study is as follows:

1. Results greatly improved with calibration, in fact, as high as a factor of five.

2. Models consistently obtained better results when used with certain types of applications.

The IITRI study was significant because it analyzed the results of seven cost models (PRICE-S, two variants of COCOMO, System-3, SPQR/20, SASET, SoftCost-Ada) to eight Ada specific programs. Ada was specifically designed for and is the principal language used in military applications, and more specifically, weapons system software. Weapons system software is different then the normal corporate type of software, commonly known as Management Information System (MIS) software. The major differences between weapons system and MIS software are that weapons system software is real time and uses a high proportion of complex mathematical coding. Up to 1997, DOD mandated Ada as the required language to be used unless a waiver was approved. Lloyd Mosemann stated: The results of this study, like other studies, showed estimating accuracy improved with calibration. The best results were achieved by SEER-SEM model were accurate within 30 percent, 62 percent of the time.

IV. SLIM ESTIAMTION MODEL

SLIM Software Life-Cycle Model was developed by Larry Putnam [3]. SLIM hires the probabilistic principle called Rayleigh distribution between personnel level and time. SLIM is basically applicable for large projects exceeding 70,000 lines of code. [4].

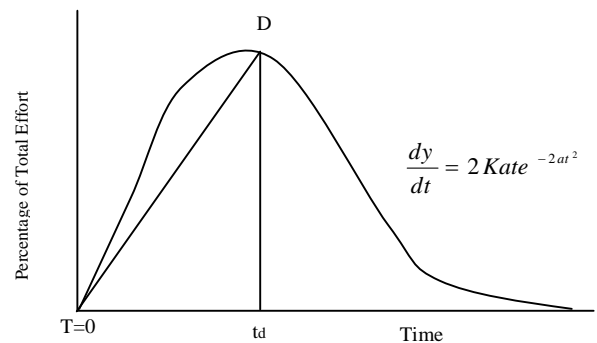

$$\frac{dy}{dt} = 2\,Kate^{-2at^2}$$

Figure 2. The Rayleigh Model

It makes use of Rayleigh curve referred from [14] as shown in figure 2 for effort prediction. This curve represents manpower measured in person per time as a function of time. It is usually expressed in personyear/ year (PY/YR). It can be expressed as

$$\frac{dy}{dt} = 2\,Kate^{-2at^2} \quad (4)$$

where,

dy/dt is the manpower utilization per unit time, " t" is the elapsed time, "a" is the parameter that affects the shape of the curve and "K" is the area under the curve. There are two important terms associated with this curve:

1) Manpower Build up given by $D_0 = K/t_d^3$

2) Productivity = Lines of Code/ Cumulative Manpower i.e. $P=S/E$ and $S=CK^{1/3}t_d^{4/3}$, where C is the technology factor which reflects the effects of various factors on productivity such as hardware constraints, program complexity, programming environment and personal experience.

The SLIM model uses two equations: the software the manpower equation and software productivity level equation The SLIM model uses Rayleigh distribution to estimate to estimate project schedule and defect rate. Two key attributes used in SLIM method are productivity Index (PI) and Manpower Buildup Index (MBI). The PI is measure of process efficiency (cost-effectiveness of assets), and the MBI determines the effects on total project effort that result from variations in the development schedule [A Probabilistic Model].

*Inputs Required:* To use the SLIM method, it is necessary to estimate system size, to determine the technology factor, and appropriate values of the manpower acceleration. Technology factor and manpower acceleration can be calculated using similar past projects. System size in terms of KDSI is to be subjectively estimated. This is a disadvantage, because of the difficulty of estimating KDSI at the beginning of a project and the dependence of the measure on the programming language.

*Completeness of Estimate*: The SLIM model provides estimates for effort, duration, and staffing information for the total life cycle and the development part of the life cycle. COCOMO I provides equations to estimate effort, duration, and handles the effect of re-using code from previously developed software. COCOMO II provides cost, effort, and schedule estimation, depending on the model used (i.e., depending on the degree of product understanding and marketplace of the project). It handles the effect of reuse, re-engineering, and maintenance adjusting the used size measures using parameters such as percentage of code modification, or percentage of design modification

*Assumptions:* SLIM assumes the Rayleigh curve distribution of staff loading. The underlying Rayleigh curve assumption does not hold for small and medium sized projects. Cost estimation is only expected to take place at the start of the design and coding, because requirement and specification engineering is not included in the model.

*Complexity:* The SLIM model's complexity is relatively low. For COCOMO the complexity increases with the level of detail of the model. For COCOMO I the increasing levels of detail and complexity are the three model types: basic, intermediate, and detailed. For COCOMO II the level of complexity increases according to the following order: Application Composition, Early Design, Post Architecture.

*Automation of Model Development:* The Putnam method is supported by a tool called SLIM (Software Life-Cycle Management). The tool incorporates an estimation of the required parameter technology factor from the description of the project. SLIM determines the minimum time to develop a given software system. Several commercial tools exist to use COCOMO models.

*Application Coverage:* SLIM aims at investigating relationships among staffing levels, schedule, and effort. The SLIM tool provides facilities to investigate trade-offs among cost drivers and the effects of uncertainty in the size estimate.

*Generalizability:* The SLIM model is claimed to be generally valid for large systems. COCOMO I was developed within a traditional development process, and was a priori not suitable for incremental development. Different development modes are distinguished (organic, semidetached, embedded). COCOMO II is adapted to feed the needs of new development practices such as development processes tailored to COTS, or reusable software availability. No empirical results are currently available regarding the investigation these capabilities.

*Comprehensiveness:* Putnam's method does not consider phase or activity work breakdown. The SLIM tool provides information in terms of the effort per major activity per month throughout development. In addition, the tool provides error estimates and feasibility analyses. As the model does not consider the requirement phase, estimation before design or coding is not possible. Both COCOMO I and II are extremely comprehensive. They provide detailed activity distributions of effort and schedule. They also include estimates for maintenance effort, and an adjustment for code re-use. COCOMO II provides prototyping effort when using the Application Composition model. The Architectural Design model involves estimation of the actual development and maintenance phase. The granularity is about the same as for COCOMO I.

V.  REVIC ESTIMATION MODEL

REVIC (REVised version of Intermediate COCOMO) is a direct descendent of COCOMO. Ourada [16] was one of the first to analyze validation, using a large Air Force database for calibration of the REVIC model. There are several key differences between REVIC and the 1981 version of COCOMO, however.

- REVIC adds an Ada development mode to the three original COCOMO modes; Organic, Semi-detached, and Embedded.

- REVIC includes Systems Engineering as a starting phase as opposed to Preliminary Design for COCOMO.

- REVIC includes Development, Test, and Evaluation as the ending phase, as opposed to COCOMO ending with Integration and Test.

- The REVIC basic coefficients and exponents were derived from the analysis of a database of completed DoD projects. On the average, the estimates obtained with REVIC will be greater than the comparable estimates obtained with COCOMO.

- REVIC uses PERT (Program Evaluation and Review Technique) statistical techniques to determine the lines-of-code input value, Low, high, and most probable estimates for each program component are used to calculate the effective lines-of-code and the standard deviation. The effective lines-of-code and standard deviation are then used in the estimation equations rather than the linear sum of the line-of-code estimates.

- REVIC includes more cost multipliers than COCOMO. Requirements volatility, security, management reserve, and an Ada mode are added.

## VI. SASET ESTIMATIN MODEL

SASET (Software Architecture, Sizing and Estimating Tool) is a forward chaining, rule-based expert system using a hierarchically structured knowledge database of normalized parameters to provide derived software sizing values. These values can be presented in many formats to include functionality, optimal development schedule, and man-loading charts. SASET was developed by Martin Marietta Denver Aerospace Corp. on contract to the Naval Center for Cost Analysis. To use SASET, the user must first perform a software decomposition of the system and define the functionalities associated with the given software system [22].

SASET uses a tiered approach for system decomposition; Tier 1 addresses software developmental and environmental issues. These issues include che class of the software to be developed, programming language, developmental, schedule, security, etc. Tier 1 output values represent preliminary budget and schedule multipliers. Tier II specifies the functional aspects of the software system, specifically the total lines-of-code (LOC). The total LOC estimate is then translated into a preliminary budget estimate and preliminary schedule estimate. The preliminary budget and schedule estimates are derived by applying the multipliers from Tier I to the total LOC estimate. Tier III develops the software complexity issues of the system under study. These issues include: level of system definition, system timing and criticality, documentation, etc. A complexity multiplier is then derived and used to alter the preliminary budget and schedule estimates from Tier II. The software system effort estimation is then calculated. Tier IV and V are not necessary for an effort estimation. Tier IV addresses the in-scope maintenance associated with the project.

The output of Tier IV is the monthly man-loading for the maintenance life-cycle. Tier V provides the user with a capability to perform risk analysis on the sizing, schedule and budget data. The actual mathematical expressions used in SASET are published in the User's Guide, but the Guide is very unclear as to what they mean and how to use them

## VII. COSTMODL ESTIMATION MODEL

COSTMODL (Cost MODeL) is a COCOMO based estimation model developed by the NASA Johnson Space Center. The program delivered on computer disk for COSTMODL includes several versions of the original COCOMO and a NASA developed estimation model KISS (Keep It Simple, Stupid). The KISS model will not be evaluated here, but it is very simple to understand and easy to use; however, the calibration environment is unknown. The COSTMODL model includes the basic COCOMO equations and modes, along with some modifications to include an Ada mode and other cost multipliers.

The COSTMODL as delivered includes several calibrations based upon different data sets. The user can choose one of these calibrations or enter user specified values. The model also includes a capability to perform a self-calibration. The user enters the necessary information and the model will "reverse" calculate and derive the coefficient and exponent or a coefficient only for the input environment data. The model uses the COCOMO cost multipliers and does not include more as does REVIC. This model includes all the phases of a software life cycle. PERT techniques are used to estimate the input lines-of-code in both the development and maintenance calculations

## VIII. STUDY OF EMPIRICAL MODELS

Empirical estimation models were studied for the past couple of decades, out of these studies many came with the result of accuracy and performance. Table I summaries the brief study of the most relevant empirical models. Studies are listed in chronological order. For each study, estimation methods are ranked according to their performance. A "1" indicates the best model, "2" the second best, and so on.

TABLE I. SUMMARY OF EMPIRICAL ESTIMATION STUDY

| Sl No. | Author | Regression | COCOMO | Analogy | SLIM | CART | ANN | Stepwise ANOVA | OSR | Expert Judgment | Other Methods |
|---|---|---|---|---|---|---|---|---|---|---|---|
| 1 | Luciana Q, 2009 | 1 | | | | | | | | | |
| 2 | Yeong-Seok Seo, 2009 | 1 | | | | | | | | | |
| 3 | Jianfeng Wen, 2009 | | 2 | | | | | | | | 1 |
| 4 | Petrônio L. Braga, 2007 | 2 | | | | | | | | | 1 |
| 5 | Jingzhou Li, 2007 | | 2 | 1 | | | | | | | |
| 6 | Iris Fabiana de Barcelos Tronto, 2007 | 3 | 4 | 5 | | | 2 | | | | 1 |
| 7 | Chao-Jung Hsu, 2007 | | | | | | | | | | 1 |
| 8 | Kristian M Furulund, 2007 | | | | | | | | | | 1 |
| 9 | Bilge Başkeleş, 2007 | | 2 | | | | | | | | 1 |
| 10 | Da Deng, 2007 | | 2 | | | | | | | | 1 |
| 11 | Simon, 2006 | | 2 | | | | | | | | 1 |
| 12 | Tim Menzies, 2005 | | 2 | | | | | | | | 1 |
| 13 | Bente Anda, 2005 | | 2 | 1 | | | | | | | |
| 14 | Cuauhtémoc López Martín, 2005 | | | | | | | | | | 1 |
| 15 | Parag C, 2005 | | | | | 2 | 3 | | | | 1 |
| 16 | Randy K. Smith, 2001 | | 2 | | | | | | | | 1 |
| 17 | Myrtveit, Stensrud, 1999 | 2 | | 3 | | | | | | | |
| 18 | Walkerden, Jeffery, 1999 | 2 | | 1 | | | | | | | |
| 19 | Kitchenham, 1998 | | | | | 2 | 1 | | | | |
| 20 | Finnie et al., 1997 | 2 | | 1 | | | 1 | | | | |
| 21 | Shepperd, Schofield, 1997 | 2 | | 1 | | | | | | | |
| 22 | Jorgensen, 1995 | 1 | | | | | 2 | | 1 | | |
| 23 | Srinivasan, Fischer, 1995 | 2 | 4 | | 5 | 3 | 1 | | | | |
| 24 | Bisio, Malabocchia, 1995 | | 2 | 1 | | | | | | | |
| 25 | Subramanian, Breslawski 1993 | 1 | 2 | | | | | | | | |
| 26 | Mukhopadhyay, Kerke 1992 | 1-3 | 2 | | | | | | | | |
| 27 | Mukhopadhyay et al., 1992 | 3 | 4 | 2 | | | | | | 1 | |
| 28 | Briand et al. 1992 | 2 | 3 | | | | | | 1 | | |
| 29 | Vicinanza et al., 1991 | 2 | 3 | | | | | | | 1 | |

*A. Impact of Cost Drivers*

Empirical Software estimation models mainly stands over the cost drivers and scale factors. These model reveals the problem of instability due to values of the cost drivers and scale factors, thus affects the sensitivity of the effort. Also, most of the model depends on the size of the project, a change in the size leads to the proportionate change in the effort. Miscalculations of the cost drives have even more vivid change in the result too. For example, a misjudgment in personnel capability in COCOMO or REVIC from 'very high to very low' will result in 300% increase in effort. Similarly in SEER-SEM changing security requirements from 'low' to 'high' will result in 400% increase in effort. In PRICE-S, 20% change in effort will occur due to small change in the value of the Productivity factor. All models have one or more inputs for which small changes will result in large changes in effort and, perhaps, schedule.

The input data problem is further compounded in that some inputs are difficult to obtain, especially early in a program. The size must be estimated early in a program using one or more sizing models. These models usually have not been validated for a wide range of projects. Some sensitive inputs, such as analyst and programmer capability, are subjective and often difficult to determine. Studies like one performed by Brent L. Barber, *Investigative Search of Quality Historical Software Support Cost Data and Software Support Cost-Related Data*, show that personnel parameter data are difficult to collect. **Figure 3**, extended from the *SEER-SEM User's Manual* shows

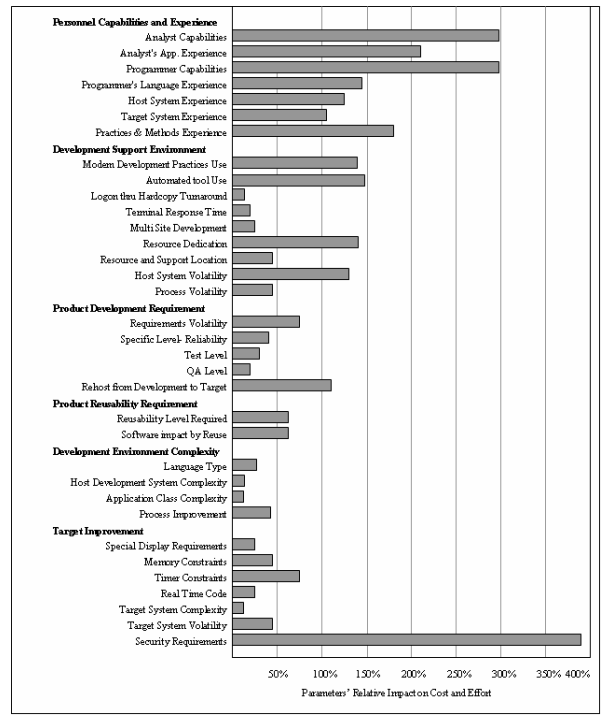

Figure 3. Relative Cost Driver Impact[32]

the relative impact on cost/effort of the different input parameters for that model. Even "objective" inputs like

Security Requirements in SEER-SEM may be difficult to confirm early in a program, and later changes may result in substantially different cost and schedule estimates. Some sensitive inputs such as the PRICE-S Productivity Factor and the SLIM PI should be calibrated from past data. If data are not available, or if consistent values of these parameters cannot be calibrated, the model's usefulness may be questionable.

## IX. ANALYSIS OF STUDY

### A. Accuracy Estimatos

The field of cost estimation suffers a lack of clarity about the interpretation of a cost estimate. Jørgensen reports different application of the term 'cost estimate' being "the most likely cost, the planned cost, the budget, the price, or, something else". Consequently there is also disagreement about how to measure the accuracy of estimates. Various indicators for accuracy – relative and absolute – have been introduced throughout the cost estimation literature such as mean squared error (MSE), absolute residuals (AR) or balanced residual error (BRE). Our literature review indicated that the most commonly used by far are the mean magnitude relative error or MMRE, and prediction within $x$ or PRED($x$). Of these two, the MMRE is the most widely used, yet both are based on the same basic value of magnitude relative error (MRE) which is defined as The first step will be to apply Conte's criteria to determine the accuracy of the calibrated and uncalibrated model. This will be achieved using the following equations.

### Conte's Criteria:

The performance of model generating continuous output can be assesses in many ways including PRED(30), MMRE, correlation etc., PRED(30) is a measure calculated from the relative error, or RE, which is the relative size of the difference between the actual and estimated value. One way to view these measures is to say that training data contains records with variables 1,2,3,……N and performance measures and additional new variables N+1, N+2,….

*MRE(Magnitude of Relative Error):* First, calculate the Magnitude of Relative Error (degree of estimating error in an individual estimate) for each data point. This step is a precedent to the next step and is also used to calculate PRED(n). Satisfactory results are indicated by a value of 25 percent or less [30].

$$MRE = |predicted - actual| / actual \quad (5)$$

*MMRE(mean magnitude of the relative error):* The mean magnitude of the relative error, or MMRE, is the average percentage of the absolute values of the relative errors over an entire data set.

$$MMRE = \frac{100}{N} \sum |predicted_i - actual_i| / actual_i \quad (6)$$

where, N = total number of estimates

*RMS (Root Mean Square):* Now, calculate the Root Mean Square (model's ability to accurately forecast the individual actual effort) for each data set. This step is a precedent to the next step only. Again, satisfactory results are indicated by a value of 25 percent or less[30].

$$RMS = \sqrt{(1/n * \sum (predicted_i - actual_i)^2)} \quad (7)$$

*RRMS(Relative Root Mean Square):* Lastly, calculate the Relative Root Mean Square (model's ability to accurately forecast the average actual effort) for each data set. According to Conte, the RRMS should have a value of 25 percent or less[30].

$$RRMS = RMS / (\sum actual / T) \quad (8)$$

*PRED(n):* A model should also be within 25 percent accuracy, 75 percent of the time [30]. To find this accuracy rate PRED(n), divide the total number of points within a data set that have an MRE = 0.25 or less (represented by k) by the total number of data points within the data set (represented by n). The equation then is: PRED(n) = k/n where n equals 0.25 [30]. In general, PRED(n) reports the average percentage of estimates that were within n percent of the actual values. Given N datasets, then

$$PRED(n) = \frac{100}{N} \sum_{i=1}^{N} \begin{cases} 1 & \text{if } MRE_i <= n/100 \\ 0 & \text{otherwise} \end{cases} \quad (9)$$

For example, PRED(30) = 50% means that half the estimates are within 30 percent of the actual.

Wilcoxon Signed-Rank Test.

The next step will be to test the estimates for bias. The Wilcoxon signed-rank test is a simple, nonparametric test that determines level of bias. A nonparametric test may be thought of as a distribution-free test; i.e. no assumptions about the distribution are made. The best results that can be achieved by the model estimates is to show no difference between the number of estimates that over estimated versus those that under estimated. The Wilcoxon signed-rank test is accomplished using the following steps [31],

1. Divide each validated subset into two groups based on whether the estimated effort was greater (T+) or less (T-) than the actual effort.

2. Sum the absolute value of the differences for the T+ and T-groups. The closer the sums of these values for each group are to each other, the lower the bias.

3. Any significant difference indicates a bias to over or under estimate.

Another performance measure of a model predicting numeric values is the correlation between predicted and actual values. Correlation ranges from +1 to -1 and a correlation of +1 means that there is a perfect positive linear relationship between variables. And can be calculates as follows

The correlation coefficient for COCOMO II is 0.6952 and the correlation coefficient for proposed model is 0.9985

$$P = \frac{\sum_i^T Predicted_i}{T}, \quad a = \frac{\sum_i^T Actual_i}{T},$$

$$S_p = \frac{\sum_i^T (Predicted_i - p)^2}{T-1}, \quad S_a = \frac{\sum_i^T (Actual_i - a)^2}{T-1},$$

$$S_{pa} = \frac{\sum_i^T (Predicted_i - p)(Actual_i - a)}{T-1},$$

$$Corr = S_{pa} / \sqrt{S_p * S_a} \quad (10)$$

All these performance measures (correlation, MMRE, and PRED) address subtly different issues. Overall, PRED measures how well an effort model performs, while MMRE measures poor performance.

A single large mistake can skew the MMREs and not effect the PREDs. Sheppard and Schofield comment that MMRE is fairly conservative with a bias against overestimate while PRED(30) will identify those prediction systems that are generally accurate but occasionally wildly inaccurate[28]

### B. Model Accuracy

There is no proof on software cost estimation models to perform consistently accurate within 25% of the cost and 75% of the time[30]. In general model fails to produce accurate result with perfect input data. The above studies have compared empirical estimation models with known input data and actual cost and schedule information, and have not found the accuracy to be scintillating. Most model were accurate within 30% of the actual cost and 57% of the time.

The Ourada study showed even worse results for SEER-SEM, SASET, and REVIC for the 28 military ground programs in an early edition of the Space and Missiles Center database. A 1981 study by Robert Thibodeau entitled *An Evaluation of Software Cost Estimating Models* showed that calibration could improve model accuracy by up to 400%. However, the average accuracy was still only 30% for an early version of SLIM and 25% for an early version of PRICE-S. PRICE-S and System-3 are within 30%, 62% of the time. An Air Force study performed by Ferens in 1983 and published in the ISPA Journal of Parametrics, concluded that no software support cost models could be shown to be accurate. The software support estimation problem is further convoluted by lack of quality software support cost data for model development, calibration, and validation. Even if models can be shown to be accurate, another effect must be considered.

Table III summarizes the parameters used and activities covered by the models discussed. Overall, model based techniques are good for budgeting, tradeoff analysis, planning and control, and investment analysis. As they are calibrated to past experience, their primary difficulty is with unprecedented situations.

TABLE II. ANALYSIS OF EMPIRICAL ESTIMATION MODELS

| Study | Model | Application Type | Validated Accuracy | | |
|---|---|---|---|---|---|
| | | | MMRE / MRE | RRMS | Pred |
| Karen Lum, 2002 | COCOMO | Flight Software | 34.4 | - | - |
| | SEER | | 140.7 | - | - |
| | COCOMO | Ground Software | 88.22 | - | - |
| | SEER | | 552.33 | - | - |
| Karen Lum, 2006 | COSEEKMO | Kind:min | 31 | - | 60(0.3) |
| | | Lang:ftn | 44 | - | 42(0.3) |
| | | Kind:max | 38 | - | 52(0.3) |
| | | All | 40 | - | 60(0.3) |
| | | Mode:org | 32 | - | 62(0.3) |
| | | Lang:mol | 36 | - | 56(0.3) |
| | | Project:Y | 22 | - | 78(0.3) |
| | | Mission Planning | 36 | - | 50(0.3) |
| | | Avioicsmonitoring | 38 | - | 53(0.3) |
| | | Mode:sd | 33 | - | 62(0.3) |
| | | Project:X | 42 | - | 42(0.3) |
| | | Fg:g | 32 | - | 65(0.3) |
| | | Center:5 | 57 | - | 43(0.3) |
| | | All | 48 | - | 43(0.3) |
| | | Mode:e | 64 | - | 42(0.3) |
| | | Cemter:2 | 22 | - | 83(0.3) |
| Gerald L Ourada, 1992 | REVIC | Aero Space | 0.373 | 0.776 | 42(0.25) |
| | SASET | | 5.95 | -0.733 | 3.5(0.25) |
| | SEER | | 3.55 | -1.696 | 10.7(0.25) |
| | Cost Model | | 0.46 | 0.53 | 29(0.25) |
| Chris F Kemour, 1987 | SLIM | ABC Software | 771.87 | - | - |
| | COCOMO | | 610.09 | - | - |
| | FP | | 102 | - | - |
| | Estimac | | 85.48 | - | - |
| | SLIM | Business:App | 772 | - | - |
| | COCOMO | | 601 | - | - |
| | FP | | 102.74 | - | - |
| | Estimac | | 85 | - | - |
| Jeremiah D Deng, 2009 | Machine Learning | Random | 0.61 | - | 0.4(30) |
| De Tran-Cao, 2007 | Cosmic | B-1 | 39 | - | 50(0.25) |

In Table II, Summary of the analysis of the study, the result of a collaborative effort of the authors, which includes author name, cost model name, application type, validated accuracy (MMRE, RRMS, Pred) is the percentage of estimates that fall within the specified prediction level of 25 or 30 percent. In this Chris F Kemour validated SLIM and obtained the result of highest MRE of 772, COCOMO obtained the result of highest

TABLE III. ACTIVITIES COVERED/FACTORS EXPLICITLY CONSIDERED BY VARIOUS MODELS[33]

| Group | Factor | SLIM | CheckPoint | Price-s | Estimacs | SEER-SEM | Select Estimator | COCOMO II |
|---|---|---|---|---|---|---|---|---|
| Size Attributes | Source Instruction | Yes | Yes | Yes | No | Yes | No | Yes |
|  | Function Points | Yes | Yes | Yes | Yes | Yes | No | Yes |
|  | OO-Related Metrics | Yes | Yes | Yes | ! | Yes | Yes | Yes |
| Program Attributes | Type/Domain | Yes | Yes | Yes | Yes | Yes | Yes | No |
|  | Complexity | Yes | Yes | Yes | Yes | Yes | Yes | Yes |
|  | Language | Yes | Yes | Yes | ! | Yes | Yes | Yes |
|  | Reuse | Yes | Yes | Yes | ! | Yes | Yes | Yes |
|  | Required Reliability | ! | ! | Yes | Yes | Yes | No | Yes |
| Computer Attributes | Resource Constraints | Yes | Yes | Yes | Yes | Yes | No | Yes |
|  | Platform Volatility | ! | ! | ! | ! | Yes | No | Yes |
| Personnel Attributes | Personnel Capability | Yes | Yes | Yes | Yes | Yes | Yes | Yes |
|  | Personnel Continuity | ! | ! | ! | ! | ! | No | Yes |
|  | Personnel Experience | Yes | Yes | Yes | Yes | Yes | No | Yes |
| Project Attributes | Tools and Techniques | Yes | Yes | Yes | Yes | Yes | Yes | Yes |
|  | Breakage | Yes | Yes | Yes | ! | Yes | Yes | Yes |
|  | Schedule Constraints | Yes | Yes | Yes | Yes | Yes | Yes | Yes |
|  | Process Maturity | Yes | Yes | ! | ! | Yes | No | Yes |
|  | Team Cohesion | ! | Yes | Yes | ! | Yes | Yes | Yes |
|  | Security Issues | ! | ! | ! | ! | Yes | No | No |
|  | Multi Site Development | ! | Yes | Yes | Yes | Yes | No | Yes |
| Activity Covered | Inception | Yes | Yes | Yes | Yes | Yes | Yes | Yes |
|  | Elaboration | Yes | Yes | Yes | Yes | Yes | Yes | Yes |
|  | Construction | Yes | Yes | Yes | Yes | Yes | Yes | Yes |
|  | Transition and Maintenance | Yes | Yes | Yes | No | Yes | No | Yes |

MRE rate of 610.09. Karen Lum has validated SEER-SEM and found to be 552.33. the validation of COSEEKMO in Mode:e has an enormous MRE of 64 evaluated by Karen Lum. Ourada study showed REVIC has MMRE of 0.373 and SASET even worse result of 5.95.

X. CONCLUSION

Based upon the background readings, this paper states that the existing models were highly credible; however, this survey found this not to be so based upon the research performed. All the models could not predict the actual against either the calibration data or validation data to any level of accuracy or consistency. Surprisingly, SEER and machine learning techniques were reliable good at predicting the effort. But however they are not accurate because all the model lies in the term prediction, prediction never comes true is proved in this estimation models. In all the models, the two key factors that influenced the estimate were project size either in terms of LOC or FP and the capabilities of the development team personnel. This paper is not convinced that no model is so sensitive to the abilities of the development team can be applied across the board to any software development effort. Finally this paper concludes that the no model is best for all situations and environment.

AUTHORS PROFILE

**Saleem Basha** is a Ph.D research scholar in the Department of Computer Science, Pondicherry University. He has obtained B.E in the field of Electrical and Electronics Engineering, Bangalore University, Bangalore, India and M.E in the field of Computer Science and Engineering, Anna University, Chennai, India. He is currently working in the area of SDLC specific effort estimation models and web service modelling systems.

**Dr. Dhavachelvan Ponnurangam** is working as Associate Professor, Department of Computer Science, Pondicherry University, India. He has obtained his M.E. and Ph.D. in the field of Computer Science and Engineering in Anna University, Chennai, India. He is having more than a decade of experience as an academician and his research areas include Software Engineering and Standards, web service computing and technologies. He has published around 75 research papers in National and International Journals and Conferences. He is collaborating and coordinating with the research groups working towards to develop the standards for Attributes Specific SDLC Models & Web Services computing and technologies.